\documentclass{article}
\usepackage[utf8]{inputenc}
\usepackage{amssymb}
\usepackage[colorlinks=true, linkcolor=blue, urlcolor=blue, citecolor=blue, plainpages=false, pdfpagelabels,linktocpage]{hyperref}
\usepackage[normalem]{ulem}

\def\newblock{\hskip .11em plus .33em minus .07em}

\begin{document}

\title{Algebraic construction of a Nambu bracket for the two-dimensional vorticity equation}

\author{M Sommer\footnote{matthias.sommer@univie.ac.at} $^{1, 3}$, K Brazda$^{1, 3}$ and M Hantel$^{2, 3}$ \\ \\
{\small $^1$ Faculty of Mathematics, University of Vienna}\\ {\small Nordbergstra\ss{}e 15, 1090 Vienna, Austria}\\
{\small $^2$ Institute of Meteorology and Geophysics, University of Vienna}\\ {\small Althanstra\ss{}e 14, 1090 Vienna, Austria}\\
{\small $^3$ Theoretical Meteorology Research Forum, University of Vienna}\\  {\small Berggasse 11, 1090 Vienna, Austria}}

\maketitle

\begin{abstract}
So far fluid mechanical Nambu brackets have mainly been given on an intuitive basis. Alternatively an algorithmic construction of
such a bracket for the two-dimensional vorticity equation is presented here. Starting from the Lie--Poisson form and its algebraic properties it is shown how the Nambu representation can be explicitly constructed as the continuum limit from the structure preserving Zeitlin discretization.
\end{abstract}



\section{Introduction}\label{intro_sec}

The Hamiltonian description has long been a matter of interest in fluid mechanics and it has enabled theoretical insight to turbulence \cite{Morrison1980, Salmon1988} and also practical applications \cite{Salmon1983, Frank2002}. In a series of articles a generalization of Hamiltonian fluid mechanics in terms of Nambu mechanics \cite{Nambu1973} has been proposed \cite{Nevir1993, Salmon2005, Bihlo2008, Nevir2009, Salazar2010}. Perspectives for the numerical analysis of fluid motion based on this formalism have also been given \cite{Salmon2005, Sommer2009, Bihlo2011}.
Not much is known however about the mathematical background of the Nambu brackets in fluid mechanics and of generic PDEs, specifically a general algorithm for finding these brackets is lacking as mentioned in \cite{Salmon2005}. Addressing this topic we present here a constructive derivation of the Nambu bracket for the example of the two-dimensional barotropic vorticity equation. Moreover we discuss the generalized Jacobi identity,
the validity of which has not been investigated thoroughly so far for fluid-mechanical Nambu brackets, or has been completely ignored. The overall result may hence be of interest given the increasing number of publications on Nambu representation in fluid mechanics.

The further part of section \ref{intro_sec} gives an overview of the state of the art, comprising the Zeitlin truncation scheme for the vorticity equation \cite{Zeitlin1991}, Nambu brackets \cite{Nambu1973}, and a general method by Bialynicki-Birula and Morrison \cite{Bialynicki1991} for their construction in the discrete case. Using the latter in section \ref{Nambu_sec} a Nambu representation of the vorticity equation is constructively derived from the underlying algebraic properties. Based on our findings we conclude with discussing the generalized Jacobi identity.

\subsection{The Zeitlin truncation scheme for the vorticity equation}

We consider the motion of an ideal, incompressible fluid on a compact domain $\Omega$. In Lagrangian view, the phase space is the cotangential bundle of the Lie group of volume preserving smooth diffeomorphisms on $\Omega$. This description can be reduced to an Eulerian one with phase space given by the dual of the Lie algebra $\mathrm{sdiff}(\Omega)$   \cite{Ebin1970}.
In two dimensions the resulting dynamical system may be expressed through the barotropic vorticity equation
\begin{equation}
\frac{\partial\zeta}{\partial t}=-J(\psi, \zeta). \label{vorticityeq}
\end{equation}
Here $\zeta$ is the vorticity of the fluid and $\psi$ the stream function related by $\zeta=\nabla^2\psi$. The Jacobi operator is defined by $J(a,b):=\frac{\partial a}{\partial x_1}\frac{\partial b}{\partial x_2}-\frac{\partial a}{\partial x_2}\frac{\partial b}{\partial x_1}$ and the velocity of the fluid is given by $\mathbf{v}=(-\frac{\partial\psi}{\partial x_2},\frac{\partial\psi}{\partial x_1})$. For simplicity all functions are assumed to be smooth.

Under suitable (e.\,g.\ periodic) boundary conditions (\ref{vorticityeq}) can be written in Lie--Poisson form
\begin{equation}
\frac{\partial\zeta}{\partial t}=\{\zeta, \mathcal{H}\}_{\textrm{\tiny{PDE}}}.
\end{equation}
Here the Lie--Poisson bracket of the barotropic vorticity equation is defined by
\begin{equation}
\{\mathcal{F}_1, \mathcal{F}_2\}_{\textrm{\tiny{PDE}}}:=\int_{\Omega}\mathrm{d}A\,\zeta J\left(\frac{\delta\mathcal{F}_1}{\delta\zeta}, \frac{\delta\mathcal{F}_2}{\delta \zeta}\right)
\label{lpbracket}
\end{equation}
for differentiable functionals $\mathcal{F}_1, \mathcal{F}_2$ of the vorticity. The Hamiltonian is given by the kinetic energy of the fluid
\begin{equation}
\mathcal{H}:=\frac{1}{2}\int_{\Omega}\mathrm{d}A\,\mathbf{v}^2=\frac{1}{2}\int_{\Omega}\mathrm{d}A\,(\nabla\psi)^2=-\frac{1}{2}\int_{\Omega}\mathrm{d}A\,\zeta\psi.\label{energy}
\end{equation}
In the following we consider a rectangular domain of length $2\pi$ in each direction with periodic boundary conditions, i.\,e.\ the torus $T^2$. Using the fact that the Lie algebra $\mathrm{sdiff}(T^2)$ can be approximated by $\mathfrak{su}(n)$ \cite{Fairlie1989, Fairlie1989b, Hoppe1989}, a spectral truncation scheme for (\ref{vorticityeq}) preserving the Lie--Poisson structure was proposed by Zeitlin \cite{Zeitlin1991}:
\begin{equation}
\frac{\partial\widehat{\zeta}_\mathbf{i}}{\partial t}=-\frac{n}{2\pi}\sum_{\mathbf{k}\in I_n}
\frac{1}{\mathbf{k}^2}\sin\left(\frac{2\pi}{n}\mathbf{i}\times\mathbf{k}\right)\widehat{\zeta}_{(\mathbf{i}+\mathbf{k})|n}\widehat{\zeta}_{-\mathbf{k}}\qquad(\mathbf{i}\in I_n).
\label{zeitlin}
\end{equation}
Here $\widehat{\zeta}_\mathbf{k}:=\frac{1}{(2\pi)^2}\int_{T^2}\mathrm{d}A\,\zeta(\mathbf{x})e^{-i\mathbf{k}\cdot\mathbf{x}}$ denote the Fourier coefficients of the vorticity. The wave vectors accounted for are located on the integer grid without the origin 
\begin{equation}
I_n:=\left\{\mathbf{i}=(i_1,i_2)\in\mathbb{Z}^2\setminus \{\mathbf{0}\}\Big|-\frac{n-1}{2}\leq i_1, i_2\leq \frac{n-1}{2}\right\},
\end{equation}
bearing a rectangular cutoff with $n^2-1$ modes in total ($n$ odd).
The norm and the vector product of the indices are determined by $\mathbf{k}^2:=k_1^2+k_2^2$ and
$\mathbf{i}\times\mathbf{k}:=i_1k_2-i_2k_1$ respectively. The vertical dash in (\ref{zeitlin}) stands for the dimension-wise modulo function onto the grid.

The truncated form of the energy as a function of the Fourier coefficients is
\begin{equation}
H:=\frac{1}{2}(2\pi)^2\sum_{\mathbf{k}\in I_n}\frac{1}{\mathbf{k}^2}\widehat{\zeta}_{\mathbf{k}}\widehat{\zeta}_{-\mathbf{k}}.
\end{equation}
This approximates the continuum form of the energy (\ref{energy}) which follows using Parseval's identity $\frac{1}{(2\pi)^2}\int_{T^2}\mathrm{d}A\,f(\mathbf{x})\overline{g(\mathbf{x})}=\sum_{\mathbf{k}\in\mathbb{Z}^2}\widehat{f}_{\mathbf{k}}\overline{\widehat{g}_{\mathbf{k}}}$ and the reality condition $\overline{\widehat{\zeta}_{\mathbf{k}}}=\widehat{\zeta}_{-\mathbf{k}}$. It can easily be checked that the truncation (\ref{zeitlin}) is of Lie--Poisson form
\begin{equation}
\frac{\partial\widehat{\zeta}_\mathbf{i}}{\partial t}=\{\widehat{\zeta}_{\mathbf{i}},H\}_{\textrm{\tiny{ODE}}}.
\end{equation}
The bracket of this discrete dynamical system is given by
\begin{eqnarray}
\{F_1,F_2\}_{\textrm{\tiny{ODE}}}&:=&\sum_{\mathbf{i,j,k}\in I_n}\alpha^{\mathbf{k}}_{\mathbf{i}\mathbf{j}}\widehat{\zeta}_{\mathbf{k}}
\frac{\partial F_1}{\partial\widehat{\zeta}_\mathbf{i}}\frac{\partial F_2}{\partial\widehat{\zeta}_\mathbf{j}}\nonumber\\
&=&-\frac{1}{(2\pi)^2}\frac{n}{2\pi}\sum_{\mathbf{i,j}\in I_n}\sin\left(\frac{2\pi}{n}\mathbf{i}\times\mathbf{j}\right)
\widehat{\zeta}_{(\mathbf{i}+\mathbf{j})|n}\frac{\partial F_1}{\partial\widehat{\zeta}_\mathbf{i}}
\frac{\partial F_2}{\partial\widehat{\zeta}_\mathbf{j}}\label{Zeitlin_LP}
\end{eqnarray}
for scalar-valued differentiable functions $F_1,F_2$ of the vorticity modes. This Lie--Poisson bracket is completely determined by the structure constants
$(\alpha_{\mathbf{ij}}^\mathbf{k})_{\mathbf{i,j,k}\in I_n}$ of the underlying Lie algebra $\mathfrak{su}(n)$:
\begin{equation}
\alpha_{\mathbf{ij}}^\mathbf{k}:=-\frac{1}{(2\pi)^2}\frac{n}{2\pi}\sin\left(\frac{2\pi}{n}\mathbf{i}\times\mathbf{j}\right)\delta_{(\mathbf{i}+\mathbf{j})|n,\mathbf{k}},
\label{zeitlinconstants}
\end{equation}
where $\delta$ denotes the Kronecker delta. In the continuum limit $n\to\infty$ they converge to the structure constants
\begin{equation}
\widetilde{\alpha}_{\mathbf{ij}}^\mathbf{k}:=-\frac{1}{(2\pi)^2}(\mathbf{i}\times\mathbf{j})\delta_{\mathbf{i}+\mathbf{j},\mathbf{k}}
\label{spectralconstants}
\end{equation}
of $\mathrm{sdiff}(T^2)$, yielding the continuum vorticity equation (\ref{vorticityeq}) in spectral form.

\subsection{Nambu brackets and their algebraic construction}

Recall that the Nambu form of a general dynamical system with prognostic variable $z$ reads \cite{Nambu1973}
\begin{equation}
\frac{\partial z}{\partial t}=\{z,\mathcal{H}_1,\mathcal{H}_2\}.
\end{equation}
In consistence with \cite{Salmon2005} we define the Nambu bracket $\{\cdot,\cdot,\cdot\}$ by the properties trilinearity, total antisymmetry, and the Leibniz rule; $\mathcal{H}_1$ and $\mathcal{H}_2$ denote generalized Hamiltonians.
Moreover a Nambu bracket defines an algebra of conserved quantities if and only if it fulfills the generalized Jacobi identity \cite{Takhtajan1994}, see subsection \ref{ssec:Jacobi}.

In \cite{Nevir1993}, a Nambu bracket for the vorticity equation has been given:
\begin{equation}
\{\mathcal{F}_1, \mathcal{F}_2, \mathcal{F}_3\}_{\textrm{\tiny{PDE}}}:=\int_\Omega\mathrm{d}A\,\frac{\delta\mathcal{F}_1}{\delta\zeta} J\left(\frac{\delta\mathcal{F}_2}{\delta \zeta}, \frac{\delta\mathcal{F}_3}{\delta\zeta}\right).\label{nambubracket}
\end{equation}
Inserting the enstrophy Casimir
\begin{equation}
\mathcal{E}:=\frac{1}{2}\int_\Omega\mathrm{d}A\,\zeta^2
\label{enstrophy}
\end{equation}
as third argument recovers the Lie--Poisson bracket (\ref{lpbracket}) (assuming suitable boundary conditions):
\begin{equation}
\{\mathcal{F}_1, \mathcal{F}_2, \mathcal{E}\}_{\textrm{\tiny{PDE}}}=\{\mathcal{F}_1, \mathcal{F}_2\}_{\textrm{\tiny{PDE}}}.
\end{equation}
Yet, in \cite{Nevir1993} neither was a claim made about the generalized Jacobi identity nor was an algorithm for deducing the Nambu bracket (\ref{nambubracket}) given.

For ODEs, such an explicit construction method from the starting point of a Lie--Poisson bracket was presented in Bialynicki-Birula and Morrison \cite{Bialynicki1991}:
Let $(\alpha{}_{ij}^k)$ be the structure constants of a semi-simple $n$-dimensional Lie algebra $\mathfrak{g}$.
If $z_k$ ($k=1,\dots, n$) denote coordinates on the dual space $\mathfrak{g}^\ast$, the Lie--Poisson bracket is given by
\begin{equation}
\{F_1, F_2\}=\sum_{i,j,k=1}^n\alpha{}_{ij}^kz_k\frac{\partial F_1}{\partial z_i}\frac{\partial F_2}{\partial z_j}
\quad \left(F_1,F_2\in\mathcal{C}^\infty(\mathfrak{g}^\ast)\right).
\end{equation}
The Killing form
$K_{ij}:=\sum_{k,l=1}^n\alpha{}_{ik}^l\alpha{}_{jl}^k$ is invertible with inverse denoted by $K^{ij}$.
A Nambu bracket can then be defined through
\begin{equation}
\{F_1,F_2,F_3\}:=\sum_{i,j,k=1}^nN_{ijk}\frac{\partial F_1}{\partial z_i}\frac{\partial F_2}{\partial z_j}\frac{\partial F_3}{\partial z_k}
\quad \left(F_1,F_2,F_3\in\mathcal{C}^\infty(\mathfrak{g}^\ast)\right)\label{bialynicki}
\end{equation}
with Nambu tensor
\begin{equation}
N_{ijk}:=\sum_{l=1}^n\alpha{}_{ij}^{{l}}K_{{lk}}.\label{bialynickitensor}
\end{equation}
By definition a Casimir $C$ of a Poisson bracket satisfies $\{C, \cdot\}=0$. The classical example in the semi-simple Lie--Poisson case is the quadratic Casimir
\begin{equation}
C:=\frac{1}{2}\sum_{i,j=1}^nK^{ij}z_iz_j\label{classicalCasimir}.
\end{equation}
Inserting it as third argument into the Nambu bracket reproduces the Lie--Poisson bracket:
\begin{equation}
\{F_1, F_2, C\}=\{F_1, F_2\}.\label{reduction}
\end{equation}
Note that the total antisymmetry of the Nambu tensor and therefore also of the bracket follows from the antisymmetry and the Jacobi identity  of the structure constants.

\section{Construction of a Nambu bracket for the discrete and continuum vorticity equation}\label{Nambu_sec}\label{Casimirs_sec}

Unfortunately the method discussed above cannot be used directly for constructing Nambu brackets for PDEs, since the Killing form is generally not defined for infinite-dimensional Lie algebras. Specifically the Killing form
\begin{equation}
\widetilde{K}_{\mathbf{ij}}=\sum_{\mathbf{k, l}\in\mathbb{Z}^2\setminus\{\mathbf{0}\}}\widetilde{\alpha}_{\mathbf{ik}}^\mathbf{l}\widetilde{\alpha}_{\mathbf{jl}}^\mathbf{k}
=-\frac{1}{(2\pi)^4}\delta_{\mathbf{i}+\mathbf{j}, \mathbf{0}}\sum_{\mathbf{k}\in\mathbb{Z}^2\setminus\{\mathbf{0}\}}(\mathbf{i}\times\mathbf{k})^2
\end{equation}
corresponding to the structure constants (\ref{spectralconstants}) of the spectral vorticity equation diverges.
The difficulties in the application of this method to hydrodynamical brackets have also been noted by \cite{Salmon2005}.
Therefore in the following we make use of the structure preserving truncation by Zeitlin in order to derive the Nambu bracket of the vorticity equation according to the method of Bialynicki-Birula and Morrison.
This comprises the computation of the Casimir and the bracket as well as the verification of their convergence in the continuum limit.

\subsection{The quadratic Casimir for the Zeitlin discretization}

We compute the Killing form corresponding to the structure constants (\ref{zeitlinconstants}) of $\mathfrak{su}(n)$:
\begin{eqnarray}
K_{\mathbf{ij}}&=&\sum_{\mathbf{k, l}\in I_n}\alpha_{\mathbf{ik}}^\mathbf{l}\alpha_{\mathbf{jl}}^\mathbf{k}\nonumber\\
&=&-\frac{1}{2}\frac{n^2}{(2\pi)^6}\delta_{(\mathbf{i}+\mathbf{j})|n, \mathbf{0}}\sum_{\mathbf{k}\in I_n}(1-\cos\left(\frac{4\pi}{n}\mathbf{i}\times\mathbf{k}\right))
\quad(\mathbf{i,j}\in I_n).
\end{eqnarray}
By the orthogonality relation
\begin{equation}
\sum_{\mathbf{k}\in I_n}\cos\left(\frac{2\pi}{n}\mathbf{k}\cdot\mathbf{l}\right)=n^2\delta_{\mathbf{l},\mathbf{0}}-1 \label{orthogonality}
\end{equation}
and since the origin is not part of the grid, we get
$\sum_{\mathbf{k}\in I_n}\cos\left(\frac{4\pi}{n}\mathbf{i}\times\mathbf{k}\right)=-1$.
Thus the Killing form and its inverse read
\begin{equation}
K_{\mathbf{ij}}=-\frac{1}{2}\frac{n^4}{(2\pi)^6}\delta_{(\mathbf{i}+\mathbf{j})|n, \mathbf{0}}\quad\mbox{and}\quad
K^{\mathbf{ij}}=-2\frac{(2\pi)^6}{n^4}\delta_{(\mathbf{i}+\mathbf{j})|n, \mathbf{0}}.
\end{equation}
Therefore, up to the normalizing constant $r:=-\frac{1}{2}(\frac{n}{2\pi})^4$, the Casimir (\ref{classicalCasimir}) is equal to the truncated form $E$ of the enstrophy
\begin{equation}
r\,C=\frac{r}{2}\sum_{\mathbf{i},\mathbf{j}\in I_n}K^\mathbf{ij}\widehat{\zeta}_\mathbf{i}\widehat{\zeta}_\mathbf{j}
=\frac{1}{2}(2\pi)^2\sum_{\mathbf{i}\in I_n}\widehat{\zeta}_{\mathbf{i}}\widehat{\zeta}_{-\mathbf{i}}=:E.\label{Casimir_bialynicki}
\end{equation}
This indeed approximates the continuum enstrophy
\begin{equation}
\mathcal{E}=\frac{1}{2}\int_{T^2}\mathrm{d}A\,\zeta^2
=\frac{1}{2}(2\pi)^2\sum_{\mathbf{k}\in\mathbb{Z}^2\setminus\{\mathbf{0}\}}\widehat{\zeta}_{\mathbf{k}}\widehat{\zeta}_{-\mathbf{k}},
\end{equation}
which shows that the algebraic Casimir property carries over to the conservation property of the dynamical PDE system.

\subsection{The Nambu bracket for the Zeitlin discretization and its continuum limit}

According to (\ref{bialynickitensor}) we define the scaled Nambu tensor
\begin{equation}
N_{\mathbf{ijk}}:=\frac{1}{r}\sum_{\mathbf{l}\in I_n}\alpha{}_{\mathbf{ij}}^{\mathbf{l}}K_{\mathbf{lk}}
=-\frac{1}{(2\pi)^4}\frac{n}{2\pi}\sin\left(\frac{2\pi}{n}\mathbf{i}\times\mathbf{j}\right)
\delta_{(\mathbf{i}+\mathbf{j}+\mathbf{k})|n,\mathbf{0}}\label{Nambutensor_Zeitlin}
\end{equation}
with associated bracket
\begin{equation}
\{F_1, F_2, F_3\}_{\textrm{\tiny{ODE}}}:=-\frac{1}{(2\pi)^4}\frac{n}{2\pi}\sum_{\mathbf{i,j,k}\in I_n}\sin\left(\frac{2\pi}{n}\mathbf{i}\times\mathbf{j}\right)\delta_{(\mathbf{i}+\mathbf{j}+\mathbf{k})|n,\mathbf{0}}
\frac{\partial F_1}{\partial\widehat{\zeta}_\mathbf{i}}
\frac{\partial F_2}{\partial\widehat{\zeta}_\mathbf{j}}
\frac{\partial F_3}{\partial \widehat{\zeta}_\mathbf{k}}\label{Nambubracket_Zeitlin}
\end{equation}
for functions $F_1, F_2, F_3$ of the vorticity modes. By construction the antisymmetry properties follow from the antisymmetry and the Jacobi identity of the underlying Lie--Poisson bracket (\ref{Zeitlin_LP}), which is recovered by inserting the truncated enstrophy:
\begin{equation}
\{F_1, F_2, E\}_{\textrm{\tiny{ODE}}}=\{F_1, F_2\}_{\textrm{\tiny{ODE}}}.
\end{equation}
In combination with the truncated energy the corresponding dynamical system
\begin{equation}
\frac{\partial\widehat{\zeta}_\mathbf{i}}{\partial t}=\{\widehat{\zeta}_{\mathbf{i}},H,E\}_{\textrm{\tiny{ODE}}}
\end{equation}
thus also represents the Zeitlin discretization, but in a form where the conservation property for the enstrophy is obvious. Of course there are $n-2$ more
independent Casimirs in the Zeitlin discretization, but the goal here is to identify the relation to the continuum Nambu bracket (\ref{nambubracket}) which employs the enstrophy.

The above discussion carries over straight to the continuum case:
Assume the functions $F_i$ on the $n$-modal system converge to the functionals $\widehat{\mathcal{F}}_i$ on the $\infty$-modal system, which by definition are equal to the functionals $\mathcal{F}_i$ of the vorticity field:
\begin{equation}
\lim_{n\rightarrow\infty}F_i((\widehat{\zeta}_{\mathbf{k}})_{\mathbf{k}\in I_n})
=\widehat{\mathcal{F}}_i((\widehat{\zeta}_{\mathbf{k}})_{\mathbf{k}\in \mathbb{Z}^2\setminus\{\mathbf{0}\}})=\mathcal{F}_i(\zeta)
\quad\quad(i=1,2,3).
\end{equation}
Then the here defined Nambu bracket (\ref{Nambubracket_Zeitlin}) converges to the bracket (\ref{nambubracket}) of the vorticity equation\footnote{It is used that the derivatives are related through $(\frac{\delta\mathcal{F}_i}{\delta\zeta})\widehat{\:}_{-\mathbf{k}}=\frac{1}{(2\pi)^2}\frac{\partial\widehat{\mathcal{F}}_i}{\partial \widehat{\zeta}_\mathbf{k}}$ ($i=1,2,3$).}:
\begin{eqnarray}
\lim_{n\rightarrow\infty}\{F_1, F_2, F_3\}_{\textrm{\tiny{ODE}}}
&=&-\frac{1}{(2\pi)^4}\sum_{\mathbf{i,j,k}\in \mathbb{Z}^2\setminus\{\mathbf{0}\}}(\mathbf{i}\times\mathbf{j})\delta_{\mathbf{i}+\mathbf{j}+\mathbf{k},\mathbf{0}}
\frac{\partial\widehat{\mathcal{F}}_1}{\partial\widehat{\zeta}_\mathbf{i}}
\frac{\partial\widehat{\mathcal{F}}_2}{\partial\widehat{\zeta}_\mathbf{j}}
\frac{\partial\widehat{\mathcal{F}}_3}{\partial \widehat{\zeta}_\mathbf{k}} \nonumber\\
&=&\{\mathcal{F}_1, \mathcal{F}_2, \mathcal{F}_3\}_{\textrm{\tiny{PDE}}}.\label{Nambubracket_PDE}
\end{eqnarray}
This shows, that the Nambu vorticity bracket discussed in \cite{Nevir1993} can be constructed from the starting point of the Zeitlin discretization in Lie--Poisson form along the ideas stated in \cite{Bialynicki1991}.
Moreover, this discussion reveals the algebraic origin of its characteristic properties. This includes the antisymmetry property and the fact that it reduces to the Lie--Poisson bracket when used with the enstrophy.

\subsection{The generalized Jacobi identity}\label{ssec:Jacobi}
A generalization of the Jacobi identity has been included in the definition of Nambu brackets by \cite{Takhtajan1994}.
While the classical form of this identity ensures that the Poisson bracket of two conserved quantities is again a conserved quantity, the generalization does the same for a Nambu bracket of three conserved quantities. For fluid mechanical systems however this issue has not yet been thoroughly discussed.
Here as well as in (application-oriented) publications, e.\,g.\ \cite{Nambu1973, Nevir1993, Salmon2005, Salazar2010, Sommer2009}, the generalized Jacobi identity is not included in the definition of a Nambu bracket. Now we assess its validity in the case of the vorticity equation.

The generalized Jacobi identity for a Nambu bracket $\{\cdot,\cdot,\cdot\}$ reads
\begin{eqnarray}
&&\{\{F_1,F_2,F_3\},F_4,F_5\}+\{F_3,\{F_1,F_2,F_4\},F_5\} \label{FI_bracket}\\
&&+\{F_3,F_4,\{F_1,F_2,F_5\}\}-\{F_1,F_2,\{F_3,F_4,F_5\}\}=0, \nonumber
\end{eqnarray}
which must hold for arbitrary functionals $F_1,F_2,F_3,F_4,F_5$.
For a totally antisymmetric Nambu tensor $N$ defined through
\begin{equation}
\{F_1,F_2,F_3\}=\sum_{i,j,k}N_{ijk}\frac{\partial F_1}{\partial z_i}\frac{\partial F_2}{\partial z_j}\frac{\partial F_3}{\partial z_k},
\end{equation}
the generalized Jacobi identity (\ref{FI_bracket}) takes the form
\begin{eqnarray}
\sum_{i,j,k,l,p,q}
\Bigg(N_{lpq}\frac{\partial(N_{ijk}\frac{\partial F_1}{\partial z_i}\frac{\partial F_2}{\partial z_j}\frac{\partial F_3}{\partial z_k})}{\partial z_l}
\frac{\partial F_4}{\partial z_p}\frac{\partial F_5}{\partial z_q}
+ N_{lpq}\frac{\partial F_3}{\partial z_l}
\frac{\partial(N_{ijk}\frac{\partial F_1}{\partial z_i}\frac{\partial F_2}{\partial z_j}\frac{\partial F_4}{\partial z_k})}{\partial z_p}
\frac{\partial F_5}{\partial z_q}
&&\nonumber\\
+N_{lpq}\frac{\partial F_3}{\partial z_l}\frac{\partial F_4}{\partial z_p}
\frac{\partial(N_{ijk}\frac{\partial F_1}{\partial z_i}\frac{\partial F_2}{\partial z_j}\frac{\partial F_5}{\partial z_k})}{\partial z_q}
- N_{ijk}\frac{\partial F_1}{\partial z_i}\frac{\partial F_2}{\partial z_j}
\frac{\partial(N_{lpq}\frac{\partial F_3}{\partial z_l}\frac{\partial F_4}{\partial z_p}\frac{\partial F_5}{\partial z_q})}{\partial z_k}\Bigg)=0.&&
\end{eqnarray}
Terms involving second-order derivatives of $F_3,F_4,F_5$ cancel and, if $N$ is independent of $z$, by the total antisymmetry of $N$ this equation reduces to
\begin{eqnarray}
&&\sum_{i,j,k,l,p,q}\left(N_{ijk}N_{lpq}+N_{ijq}N_{lkp}+N_{ijp}N_{lqk}\right)\nonumber\\
&&\phantom{\sum_{i,j,k,l,p,q}}
\left(\frac{\partial^2 F_1}{\partial z_l\partial z_i}\frac{\partial F_2}{\partial z_j}
+\frac{\partial F_1}{\partial z_i}\frac{\partial^2 F_2}{\partial z_l\partial z_j}\right)
\frac{\partial F_3}{\partial z_k}\frac{\partial F_4}{\partial z_p}\frac{\partial F_5}{\partial z_q}=0.
\end{eqnarray}
Thus in the case of a constant Nambu tensor, the generalized Jacobi identity (\ref{FI_bracket}) is equivalent to the identity
\begin{equation}
N_{{ijk}}N_{{lpq}}+N_{{ijq}}N_{{lkp}}+N_{{ijp}}N_{{lqk}}=0\label{FI_tensor}
\end{equation}
for all $i, j, k, l, p, q$ in the finite or infinite integer range.

However, as remarked in \cite{Sahoo1993}, the generalized Jacobi identity is not necessarily fulfilled for the Nambu tensor constructed with the method of Bialynicki-Birula and Morrison.
A simple counterexample for the discrete bracket (\ref{Nambubracket_Zeitlin}) as well as for the continuum bracket (\ref{Nambubracket_PDE}) is given here:
With the wave vectors
$\mathbf{i}=(1, 0)$, $\mathbf{j}=(0, 1)$, $ \mathbf{k}=(-1, -1)$, $ \mathbf{l}=(1, 0)$, $ \mathbf{p}=(-1, 1)$, $ \mathbf{q}=(0, -1)$,
we get in both cases for the first term in (\ref{FI_tensor}) a non-vanishing value and for the others zero. Consequently, none of these Nambu brackets fulfills the generalized Jacobi identity.

\section{Conclusion}

The presented article contains a discussion on the interpretation of the vorticity equation Nambu bracket in terms of its Lie--Poisson bracket.
The algebraic Casimir of the Zeitlin discretization was computed explicitly and shown to be proportional to the truncated enstrophy. It was further demonstrated how the Nambu bracket with its characteristic properties arises naturally from the Lie--Poisson form of the vorticity equation.
The existence of a Nambu representation for the vorticity equation could be traced back to the fact that a structure preserving truncation with semi-simple Lie algebra and enstrophy as classical Casimir exists. Without contradicting this result the generalized Jacobi identity was shown not to apply.

It would be interesting to generalize the construction method for the use with vorticity Casimirs of higher degree instead of the enstrophy, with potential perspectives in numerical applications. Furthermore, while for geophysical fluid dynamics the here treated two-dimensional vorticity equation is already of interest, a future discussion could also include more sophisticated fluid mechanical systems, such as e.\,g.\ the three-dimensional incompressible or shallow-water equations. However, since no structure preserving discretizations exist in these cases \cite{Zeitlin1991}, this is presumably more challenging. It may still be hoped that the presented method can shed light on an interesting facet of the formulation of fluid mechanics.

\section*{Acknowledgments}
We thank A.~Bihlo and R.~Popovych for valuable discussions and the three referees for their comments which helped to improve the manuscript.
This work has been funded by the Austrian Science Fund (FWF) as project P21335.

\bibliographystyle{unsrt}

\end{document}